\documentclass[letterpaper, 11pt]{article}
\usepackage{float}
\usepackage[section]{placeins}
\usepackage{authblk}
\usepackage{amsmath}
\usepackage{graphicx}
\usepackage{color}
\usepackage{cancel}
\usepackage{amsfonts}
\usepackage{amssymb}
\usepackage{booktabs}
\usepackage{subfigure}
\usepackage{cite}
\usepackage[font={footnotesize,it}]{caption}

\title{Wang and Yau's Quasi-Local Energy for an Extreme Kerr Spacetime}

\author[1]{Warner A. Miller}
\author[1]{Shannon Ray}
\author[2]{Mu-Tao Wang}
\author[3]{\\Shing-Tung Yau}
\affil[1]{Department of Physics, Florida Atlantic University, Boca Raton FL 33431, USA}
\affil[2]{Department of Mathematics, Harvard University, Cambridge, MA 02138, USA}
\affil[3]{Department of Mathematics, Columbia University, New York, NY 10027, USA.}

\begin{document}

\maketitle

\begin{abstract}
There exist constant radial surfaces, $\mathcal{S}$, that may not be globally embeddable in $\mathbb{R}^3$ for Kerr spacetimes with $a>\sqrt{3}M/2$.  To compute the Brown and York (B-Y) quasi-local energy (QLE), one must isometrically embed $\mathcal{S}$ into $\mathbb{R}^3$.  On the other hand, the Wang and Yau (W-Y) QLE embeds $\mathcal{S}$ into Minkowski space.  In this paper, we examine the W-Y QLE for surfaces that may or may not be globally embeddable in $\mathbb{R}^3$.  We show that their energy functional, $E[\tau]$, has a critical point at $\tau=0$ for all constant radial surfaces in $t=constant$ hypersurfaces using Boyer-Lindquist coordinates.  For $\tau=0$, the W-Y QLE reduces to the B-Y QLE.  To examine the W-Y QLE in these cases,  we write the functional explicitly in terms of $\tau$ under the assumption that $\tau$ is only a function of $\theta$. We then use a Fourier expansion of $\tau\left(\theta\right)$ to explore the values of $E[\tau\left(\theta\right)]$ in the space of coefficients.  From our analysis, we discovered an open region of complex values for $E[\tau\left(\theta\right)]$.   We also study the physical properties of the smallest real value of $E[\tau\left(\theta\right)]$, which lies on the boundary separating real and complex energies. 
\end{abstract}

\newpage
\section{Introduction}
\label{sec:intro}   
\begin{figure}[h]
\centering
\includegraphics[width=2in]{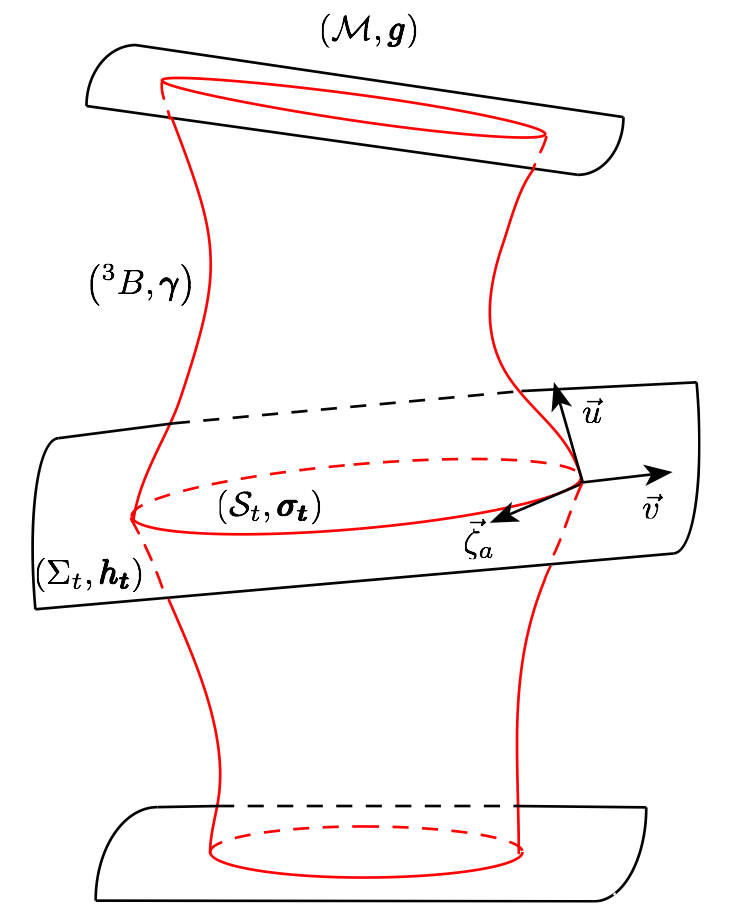}
\caption
{This figure represents a 3+1 split of a spacetime manifold with 4-metric $\left(\mathcal{M},\pmb{g}\right)$.  Here we have suppressed one spatial dimension.  The manifold  $\mathcal{M}$ is foliated by a family of spacelike hypersurfaces with 3-metric $\left(\Sigma_t,\pmb{h_t}\right)$.  Each hypersurface has a volume bounded by a simply connected spacelike surface with 2-metric $\left(\mathcal{S}_t,\pmb{\sigma_t}\right)$.  There are two spacelike  boundaries and one timelike boundary of $\mathcal{M}$.   The spacelike boundaries are the initial and final hypersurfaces $\Sigma_0$ and $\Sigma_f$ of the foliation. The timelike boundary $\left(^{3}B,\pmb{\gamma}\right)$ is a three dimensional timelike cylindrical surface that is a product of 2-surfaces $\mathcal{S}_t$ embedded in $\Sigma_t$ and the world lines of Eulerian observers.  The vector $\vec{u}$ is the timelike normal vector to $\Sigma_t$ and is tangent to $^{3}\!B$.  Vector $\vec{v}$ is orthogonal to $\mathcal{S}_t$ and $^{3}\!B$ but tangent to $\Sigma_t$.  The vectors $\vec{\zeta}_a$ span the tangent space of $\mathcal{S}_t$ and are tangent to both $\Sigma_t$ and $^{3}\!B$.}
\label{fig:central}
\end{figure}

It is not possible to define a local measure of the gravitational energy associated with the curvature of spacetime due to the equivalence principle of general relativity.  However, it is possible to define a quasi-local energy (QLE) density with respect to a field of observers $\vec{t}$ and a 2-surface $\mathcal{S}$ bounding some 3-volume in a spacetime manifold $\mathcal{M}$.  In 1993, Brown and York (B-Y) gave a natural method for devising such an energy using a Hamilton-Jacobi approach~\cite{BY:QLE}.  To understand their expression for QLE, we first introduce Fig.~\ref{fig:central}, which includes notations for all submanifolds of $\mathcal{M}$ and their respective metrics.  It also includes the notations for the normal and tangent vectors defined in $\mathcal{M}$.
Looking at equation 4.5 of~\cite{BY:QLE}, the B-Y QLE is defined as 
\begin{equation}
\label{eq:qleBY}
E=\underbrace{-\frac{1}{8\pi}\int_{\mathcal{S}_t}{[N k - N^\mu v^\nu \left(K_{\mu \nu} - K g_{\mu \nu}\right)]\sqrt{\sigma_t}\ dx^2}}_{physical\  space\  energy} - 
\underbrace{E^0}_{reference\ energy}
\end{equation} 
where $k$ is the mean curvature of $\mathcal{S}_t$ embedded in the spacelike hypersurface $\Sigma_t$, $\pmb K$ is the extrinsic curvature tensor of $\Sigma_t$ embedded in $\mathcal{M}$, $K$ is the trace of $\pmb K$ and $E^0$ is the reference energy that emerges from the freedom to choose the zero point energy in any Hamilton-Jacobi formulation. The lapse and shift are given by $N$ and $\vec{N}$, respectively.  Hawking and Horowitz proposed a similar definition of QLE in 1996~\cite{HH:QLE}.  One choice for $E^0$ suggested by B-Y involves isometrically embedding $\mathcal{S}_t$ in some flat reference space and computing the corresponding reference energy.  This gives 
\begin{equation}
\label{eq:ref}
E^0=-\frac{1}{8\pi}\int_{\mathcal{S}_t}{\left[N k_0 - N^\mu v_0^\nu \left(\left(K_0\right)_{\mu \nu}-K_0 \eta_{\mu \nu}\right)\right]\sqrt{\sigma_t}\ dx^2} 
\end{equation} 
where the $N$ and $\vec{N}$ are the same as Eq.~\ref{eq:qleBY} and $\pmb{\eta}$ is the metric of the flat space.  Their reason for choosing the reference space to be flat is one would expect the QLE to be zero for a flat spacetime. 

Given $\mathcal{S}_t$ defined in a maximal hypersurface of a stationary spacetime, B-Y suggested that one uses the Eulerian observers defined by $\vec{t}=\vec{u}$ as their observers and $\mathbb{R}^3$ as their reference space.  Using these suggestions, the B-Y QLE reduces to
\begin{equation}
\label{eq:simpBY}
E_{BY} = \frac{1}{8\pi}\int_{\mathcal{S}_t}{\left(k-k_0\right)\sqrt{\sigma_t}\ dx^2}.
\end{equation} 
The surface isometric embedding theorem (proposed by Weyl and proved independently by Nirenberg~\cite{Nirenberg:WeylProof} and Pogorelov~\cite{Pog:Reg}) states that a closed surface with
a Riemannian metric of positive Gaussian curvature can be uniquely isometrically embedded into $\mathbb{R}^3$.

\begin{figure}[h]
\centering
\includegraphics[width=3.5in]{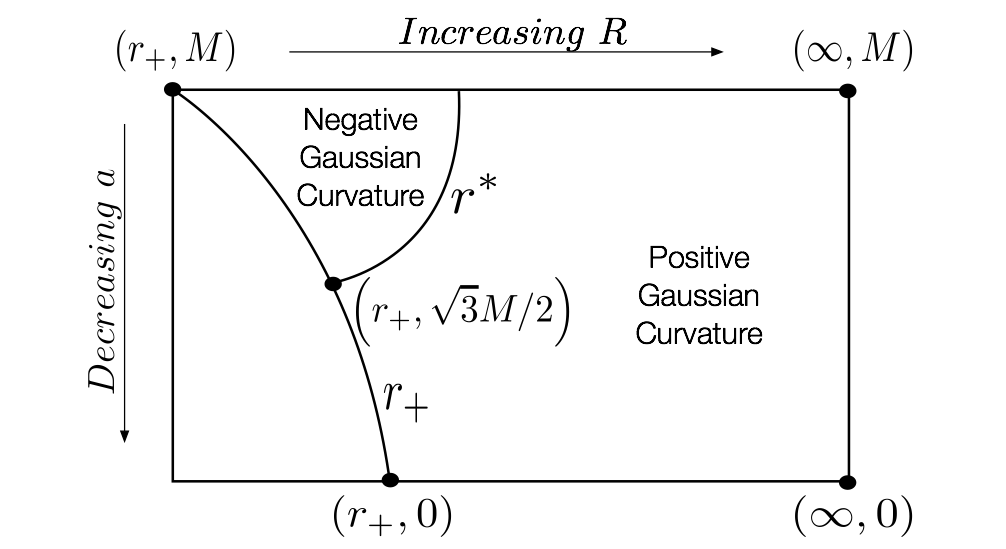}
\caption{In this figure we have the domain of QLE for Kerr spacetimes as a function of radius and angular momentum per unit mass.  Here, the curve $r_+$ represents the event horizon, while $r^*$ is the curve separating surfaces with strictly positive Gaussian curvature from those with regions of negative Gaussian curvature.}
\label{fig:demarc}
\end{figure}

In 1994, Martinez analyzed Eq.~\ref{eq:simpBY} for Kerr spacetimes using a small angular momentum approximation~\cite{Martinez:physrev}.  With this approximation, Martinez found that the B-Y QLE at the event horizon is given by
\begin{equation}
\label{eq:irr}
E=2M_{ir} = \sqrt{\left(M+\sqrt{M^2-a^2}\right)^2+a^2} 
\end{equation} 
where $M_{ir}$ is the irreducible mass, $a$ is the angular momentum per unit mass and $M$ is the mass of the black hole.  In 1973, Larry Smarr showed that the event horizon of a Kerr black hole with $a>\sqrt{3}M/2$ has a region centered at the poles with negative Gaussian curvature~\cite{Smarr:1973}.  Since the Gaussian curvature is not positive everywhere, the theorem of Nirenberg and Pogorelov is not applicable. Thus an isometric embedding into $\mathbb{R}^3$ may not exist at all, and an existing isometric embedding may not be unique. This implies that the B-Y QLE energy is not well defined at the event horizon for spacetimes with large angular momentum. See Appendix ~\ref{sec:iso} for a discussion on surface isometric embeddings. The existence of negative Gaussian curvature creates a demarcation between constant radial surfaces for which Eq.~\ref{eq:simpBY} is well defined everywhere and those where it is only partially defined.  This demarcation is illustrated in Fig.~\ref{fig:demarc}.   
One can explicitly write $r^*$ for a constant radial surface by finding the root of its Gaussian curvature at the poles.  We begin the derivation of $r^*$ by first introducing the metric of the constant radial surface in Kerr.

The line element of Kerr in Boyer-Lindquist coordinates is given by
\begin{equation}
\label{eq:kle}
dl^2_{\mathcal{M}} = g_{tt} dt^2 + 2 g_{t\phi} dt d\phi + g_{rr} dr^2 + g_{\theta\theta} d\theta^2 + g_{\phi\phi} d\phi^2
\end{equation}
where
\begin{eqnarray}
\label{eq:gtt}
g_{tt}      & = & -\left( 1- \frac{2M r}{\Xi} \right),\\
\label{eq:gtp}
g_{t\phi} & = & - \frac{2 M r}{\Xi}\, a \sin^2{\theta},\\
g_{rr}     & = &  \frac{\Xi}{\Delta},\\
g_{\theta\theta} & = & \Xi\ \ \hbox{and}\\
g_{\phi\phi}       & = & \left( r^2 + a^2 \left(1+\frac{ 2  M r \sin^2{\theta}}{\Xi}\right)\,  \right) \sin^2{\theta}
\end{eqnarray}
are the non-zero components of the Kerr metric.  The definitions of $\Xi$ and $\Delta$ are
\begin{eqnarray}
\Xi & := &  r^2+ a^2 \cos^2{\theta}\ \ \hbox{and}\\
\Delta & := & r^2-2 M r+a^2. 
\end{eqnarray}
The 2-surface $\mathcal{S}_t$ for which quasi-local energy is computed is defined in a $t=constant$ hypersurface $\Sigma$ with constant radius $R$.  The choice of $t$ is inconsequential since the spacetime is stationary.  For this reason, we drop the subindex $t$ from subsequent notation.  Inserting $dt=dr=0$ and $r=R$ in Eq.~\ref{eq:kle} gives the line element of $\mathcal{S}$ as
 \begin{equation}
 \label{eq:Smetric}
 dl_{\mathcal{S}}^2 = \underbrace{\left(R^2+a^2 \cos^2{\theta}\right)}_{\sigma_{\theta\theta}}\,  d\theta^2 +  
             \underbrace{\left( R^2+a^2+\frac{ 2 R a^2 M \sin^2{\theta}}{\Xi}  \right) \sin^2{\theta}}_{\sigma_{\phi\phi}}\, d\phi^2
 \end{equation}
where $\sigma_{\theta \theta}$ and $\sigma_{\phi \phi}$ are the non-zero components of the induced metric $\pmb{\sigma}$ on $\mathcal{S}$.  The Gaussian curvature of $\mathcal{S}$ is given by
\begin{equation}
\label{eq:gauss}
\mathcal{K} = \frac{\sigma_{\theta \theta}\sigma^2_{\phi \phi,\theta}+\sigma_{\phi \phi}\left(\sigma_{\theta \theta,\theta}\sigma_{\phi \phi,\theta}-2\sigma_{\theta \theta}\sigma_{\phi \phi,\theta \theta}\right)}{4\sigma^2_{\theta \theta}\sigma^2_{\phi \phi}}.
\end{equation}
Solving for the root of Eq.~\ref{eq:gauss} at $\theta=0$ gives
\begin{equation}
\label{eq:rstar}
r^*\left(a,M\right)=\frac{-3^{1/3}a^2+\Gamma^{2/3}}{3^{2/3}\Gamma^{1/3}}
\end{equation}
where
\begin{equation}
\Gamma=27 M a^2+a\sqrt{3}\sqrt{243 M^2 + a^4}.
\end{equation}
This is the only non-zero real root of the Gaussian curvature at the poles.  Recently, Yu and Liu studied QLE for $r^*<R<\sqrt{3}a$ with unrestricted angular momentum~\cite{CL:2017}.  Their analysis remains in the regime of strictly positive Gaussian curvature.  

In this paper, we study the Wang and Yau (W-Y) QLE for constant radial surfaces in Kerr with $r_+<R<r^*$.   In the W-Y approach, one embeds $\mathcal{S}$ into Minkowski space $\mathbb{R}^{3,1}$ instead of $\mathbb{R}^3$.  Because $\mathcal{S}$ is a co-dimension 2 surface with respect to $\mathbb{R}^{3,1}$, the isometric embedding equations are underdetermined thus giving infinitely many embeddings. To solve this problem, W-Y introduced the scalar field $\tau$ on $\mathcal{S}$, which determines a unique embedding into $\mathbb{R}^{3,1}$ given a choice of $\tau$.  Choosing $\tau$ also chooses a unique field of observers on $\mathcal{S}$.  Using the W-Y approach, Eq.~\ref{eq:qleBY} is redefined as
\begin{equation}
\label{eq:WYQLE}
E[\tau]=\underbrace{\frac{1}{8\pi}\int{\left(-\bar{k}\sqrt{1+|\nabla \tau|^2} + \langle \nabla \tau\ |\ \nabla \vec{\bar{v}}\ |\ \vec{\bar{u}}\rangle\right)\sqrt{\sigma}\ dx^2} }_{physical\ space\ energy} - \underbrace{\frac{1}{8\pi}\int{\hat{k} \sqrt{\hat{\sigma}}\ dx^2}}_{reference\ energy}
\end{equation}
where $\bar{k}$ is the mean curvature of $\mathcal{S}$ embedded in $\mathcal{M}$ with respect to the spacelike normal $\vec{\bar{v}}$ and $\hat{k}$ is the mean curvature of the convex shadow $\hat{\mathcal{S}}$ embedded in $\mathbb{R}^3$.  The 2-metric of the convex shadow is written as $\pmb{\hat{\sigma}}$.  All necessary information for this paper regarding the normal basis $\{\vec{\bar{u}}, \vec{\bar{v}}\}$ and the convex shadow is contained in appendix~\ref{sec:app}; they are also defined in~\cite{WY:comm,WY:physrev}.  The purpose of the appendix is to give the reader a self contained explanation for the physical motivations behind the W-Y formalism.  The W-Y QLE is defined as the minimum of Eq.~\ref{eq:WYQLE} with respect $\tau$, which is equivalent to minimizing with respect to all possible observer fields.  

We are unaware of any research that explores QLE near the event horizon for extreme Kerr spacetimes using a Hamilton-Jacobi approach.  Given the generalization of the B-Y QLE by W-Y, we believe their definition is a good starting point to explore this area of research.  It can be shown for a Kerr spacetime that a critical point of the Eq.~\ref{eq:WYQLE} is found at $\tau=0$ regardless of the value of $R$. Given $\tau=0$, the W-Y QLE functional reduces to Eq.~\ref{eq:simpBY}.  To gleam some insight on the behavior of $E[\tau]$ in this region, we explore the W-Y QLE using numerical techniques.  We restrict $\tau$ to only a function of $\theta$ to simplify the W-Y QLE functional and make the calculation more tractable.  Given this restriction on $\tau$, the main results of this analysis are the following: (1) there exists a boundary separating admissible real energies from inadmissible complex energies and the minimum real value, $E_{min}$, of $E[\tau\left(\theta\right)]$ lies on this boundary, (2) $\tau=0$, which is a critical point of the W-Y QLE functional for constant radial surfaces with $R<r^*$, is not admissible within their formalism and (3) the physical behavior of $E_{min}$ disagrees with the behavior one would expect from the analysis of Martinez.  

We structure the paper in the following way. In Sec.~\ref{sec:QLEintau} we write the W-Y QLE functional in terms of $\tau$.  In Sec.~\ref{sec:zeromin} we show that $\tau=0$ is a critical point of the W-Y QLE functional for Kerr regardless of the the value of $R$.  In Sec.~\ref{sec:numerical} we present our numerical analysis.  Finally, in Sec.~\ref{sec:conclusion} we have further discussions and conclusions.
\section{Expressing the W-Y QLE in terms of $\tau$}
\label{sec:QLEintau}
The purpose of this section is to write Eq.~\ref{eq:WYQLE} explicitly in terms of $\tau$ for constant radial surfaces.  This will be used in Sec.~\ref{sec:numerical} for our numerical analysis. To this end, we separate this section into two subsections.  The first derives the physical energy in terms of $\tau$, while the second derives the reference energy in terms of $\tau$.  Before we continue with our derivations,  we must define the mean curvature vector $\vec{H}$.  

Let $\vec{X}\left(\eta^a\right)$ represent the spacetime coordinates of $\mathcal{S}$ embedded in $\mathcal{M}$ where $\eta^a=\{\theta,\phi\}$ are the Boyer-Lindquist coordinates of $\mathcal{S}$.  At each point $p\in\mathcal{S}$ there also exists a spacelike tangent plane $\mathcal{T}_s\left(p\right)$ that is spanned by an orthogonal basis made of spacelike tangent vectors $\vec{\zeta}_a=\frac{\partial\vec{X}}{\partial\eta^a}$.  Given an arbitrary normal basis $\{\vec{u},\vec{v}\}$ on $\mathcal{S}$, the mean curvature vector can be written as
\begin{equation}
\vec{H}= H_{\vec{u}}\ \vec{u} + H_{\vec{v}}\ \vec{v} 
\end{equation}
where $H_{\vec{u}}$ is the fractional rate of expansion of $\mathcal{S}$ along the timelike normal $\vec{u}$ and is given by
\begin{equation}
\label{eq:ht}
H_{\vec{u}} = \sigma^{ab} \langle\vec{\zeta}_a\ |\ \nabla \vec{u}\ |\ \vec{\zeta}_b\rangle,
\end{equation}
and $H_{\vec{v}}$ is the fractional rate of expansion of $\mathcal{S}$ along the spacelike normal $\vec{v}$ and is given by
\begin{equation}
\label{eq:hs}
H_{\vec{v}} = \sigma^{ab} \langle\vec{\zeta}_a\ |\ \nabla \vec{v}\ |\ \vec{\zeta}_b\rangle.
\end{equation}
The covariant derivative is taken with respect to the Kerr metric $\pmb{g}$ for both $H_{\vec{u}}$ and $H_{\vec{v}}$.  The mean curvature vector is the direction of maximal expansion of $\mathcal{S}$ in $\mathcal{M}$ and is independent of the normal basis in which it is computed. 
\subsection{The physical contribution to the W-Y QLE in terms of $\tau$}
\label{sec:QLEphys}
In this subsection we follow the prescription given in~\cite{WY:comm} to compute the physical portion of QLE.  This is done in three steps:
\begin{enumerate}
\item Compute the normal basis $\{\vec{u'},\vec{v'}\}$ of $\mathcal{S}$ that satisfies
\begin{equation}
\label{eq:nonprefbasis}
\vec{v'} = \frac{\vec{H}}{|\vec{H}|}.
\end{equation}
\item Transform $\{\vec{u'},\vec{v'}\}$ to $\{\vec{\bar{u}},\vec{\bar{v}}\}$ using
\begin{eqnarray}
\label{eq:utilde}
\vec{\bar{u}}&=&\vec{u'}\cosh{\alpha}+\vec{v'}\sinh{\alpha} \\
\label{eq:vtilde}
\vec{\bar{v}}&=&\vec{u'}\sinh{\alpha}+\vec{v'}\cosh{\alpha}
\end{eqnarray}
where $\alpha$ is the hyperbolic angle that minimizes the physical energy in Eq.~\ref{eq:WYQLE} and is given by
\begin{equation}
\label{eq:sinh}
\sinh{\alpha} = \frac{-\Delta \tau}{|\vec{H}|\sqrt{1+|\nabla \tau|^2}}.
\end{equation}
\item Use $\{\vec{\bar{u}},\vec{\bar{v}}\}$ to express the physical energy in terms of $\tau$.
\end{enumerate}
We will refer to $\{\vec{u'},\vec{v'}\}$ and $\{\vec{\bar{u}},\vec{\bar{v}}\}$ as the non-preferred and preferred normals, respectively. 

For step 1, we begin with the non-preferred normal basis
\begin{equation}
\label{vprime}
\vec{v'}=\{0,\frac{1}{\sqrt{g_{rr}}},0,0\}
\end{equation}
and
\begin{equation}
\label{uprime}
\vec{u'} = \beta\{-\frac{g_{\phi \phi}}{g_{t\phi}},0,0,1\}
\end{equation}
where
\begin{equation}
\beta=\frac{1}{\sqrt{g_{\phi \phi}\left(1-\frac{g_{\phi \phi}g_{tt}}{g^2_{t\phi}}\right)}}.
\end{equation}
Here, $\vec{u'}$ is the timelike normal of $\Sigma$ restricted to $\mathcal{S}$.  Since $\Sigma$ is a maximal hypersurface of Kerr and $\mathcal{S} \subset \Sigma$, writing the mean curvature vector in terms of $\{\vec{u'},\vec{v'}\}$ gives
\begin{equation}
\label{eq:H}
\vec{H} = \cancelto{0}{H_{\vec{u'}}}\ \ \ \vec{u'} + H_{\vec{v'}}\ \vec{v'},
\end{equation}
which satisfies Eq.~\ref{eq:nonprefbasis} and completes step 1.  For step 2 we use $\cosh^2{\alpha}-\sinh^2{\alpha}=1$ to write
\begin{equation}
\label{eq:cosh}
\cosh{\alpha} = \sqrt{1+\frac{\left(\Delta \tau\right)^2}{|\vec{H}|^2 \left(1+|\nabla \tau|^2\right)}}.
\end{equation}
Inserting $\sinh{\alpha}$ and $\cosh{\alpha}$ into Eqs.~\ref{eq:utilde} and~\ref{eq:vtilde} to transform from $\{\vec{u'},\vec{v'}\}$ to $\{\vec{\bar{u}},\vec{\bar{v}}\}$ completes step 2.

To complete step 3, we begin by inserting $\vec{\bar{v}}$ into  Eq.~\ref{eq:hs} to compute $\bar{k}$, which gives
\begin{equation}
\label{eq:Hmag}
\bar{k}=-\sqrt{g_{rr}} \cosh{\bar{\alpha}}\left(\frac{\Gamma^r_{\theta \theta}}{\sigma_{\theta \theta}} + \frac{\Gamma^r_{\phi \phi}}{\sigma_{\phi \phi}}\right).
\end{equation} 
Next we insert $\bar{k}$ into the first term of the physical space energy giving
\begin{equation}
\label{eq:term1}
\bar{k} \sqrt{1+|\nabla \tau|^2} = -\sqrt{g_{rr}\left(\frac{\Gamma^r_{\theta \theta}}{\sigma_{\theta \theta}} + \frac{\Gamma^r_{\phi \phi}}{\sigma_{\phi \phi}}\right)^2\left(1+|\nabla \tau|^2\right)+\Delta \tau^2}
\end{equation}
where 
\begin{equation}
|\nabla \tau|^2=\frac{\tau^2_{,\theta}}{\sigma_{\theta\theta}}
\end{equation}
and
\begin{eqnarray}
\Delta \tau &=& \sigma^{ab}\nabla_a \nabla_b \tau\\
\label{eq:laptau}
&=& \frac{1}{\sigma_{\theta \theta}}\left(\tau_{,\theta \theta} - \Gamma^\theta_{\theta \theta} \tau_{,\theta} \right) - \frac{\Gamma^\theta_{\phi \phi}\tau_{,\theta}}{\sigma_{\phi \phi}}.
\end{eqnarray}
For the second term of the physical space energy, we map $\nabla \tau$ from $\mathcal{S}$ to $\mathcal{M}$ using
\begin{equation}
\label{eq:4dtau}
\nabla \tau = \sigma^{ab}\tau_{,a} \vec{\zeta}_{b}.
\end{equation}
Inserting $\vec{\bar{u}}$ and $\vec{\bar{v}}$ into the second term of the physical space energy gives
\begin{equation}
\label{eq:term2}
\langle \nabla \tau\ |\ \nabla \vec{\bar{v}}\ |\ \vec{\bar{u}}\rangle = \frac{\tau_{,\theta}}{\sigma_{\theta \theta}}\left(\left(\bar{v}_{t,\theta}-\Gamma^t_{\theta t}\bar{v}_t\right)\bar{u}^t + \left(\bar{v}_{r,\theta}-\Gamma^r_{\theta r}\bar{v}_r\right)\bar{u}^r-\Gamma^t_{\theta \phi}\bar{v}_t\bar{u}^\phi\right).
\end{equation}
Combining Eq.~\ref{eq:term1} and \ref{eq:term2}  and integrating over $\mathcal{S}$ gives the physical contribution to the W-Y QLE in terms of derivatives of $\tau$ and completes step 3.

\subsection{The reference contribution to the W-Y QLE in terms of $\tau$}
\label{sec:QLEreference}
In section 3 of~\cite{WY:comm}, it was shown that 
\begin{equation}
E^0=-\int_{\mathcal{S}}{\left[N k_0 - N^\mu v_0^\nu \left(\left(K_0\right)_{\mu \nu}-K_0 \eta_{\mu \nu}\right)\right]\sqrt{\sigma}\ dx^2} = \int{\hat{k} \sqrt{\hat{\sigma}}\ dx^2}
\end{equation}
where $\hat{k}$ is the mean curvature of the convex shadow embedded in $\mathbb{R}^3$.  Therefore, one only needs to isometrically imbed $\hat{\mathcal{S}}$ in $\mathbb{R}^3$ and integrate the mean curvature to find the reference energy.  Assuming $\tau$ is only a function of $\theta$,  the metric components of $\hat{\mathcal{S}}$ are given by 
\begin{eqnarray}
\label{eq:thetahat}
\hat{\sigma}_{\theta \theta} &=& \sigma_{\theta \theta} + \tau^2_{,\theta} \\
\label{eq:phihat}
\hat{\sigma}_{\phi \phi} &=& \sigma_{\phi \phi}.
\end{eqnarray}
Let the Cartesian coordinates of $\hat{\mathcal{S}}$ be defined as
\begin{eqnarray}
\label{eq:x}
x\left(\theta, \phi \right)&=&\rho\left(\theta\right)\cos{\phi}\\
\label{eq:y}
y\left(\theta, \phi \right)&=&\rho\left(\theta\right)\sin{\phi}\\
\label{eq:z}
z\left(\theta\right)&=&f\left(\theta\right)
\end{eqnarray}
where $\rho\left( \theta \right)$ and $f\left( \theta \right)$ are smooth real valued functions on the domain $\theta \in [0,\pi]$.  Equating the line element on $\hat{\mathcal{S}}$ with that of Euclidean space, we get
\begin{eqnarray}
\label{eq:rho}
\rho\left(\theta\right)&=&\sqrt{\sigma_{\phi \phi}} \\
\label{eq:df}
f_{,\theta}\left(\theta\right)&=& \sqrt{\sigma_{\theta \theta} - \frac{\sigma^2_{\phi \phi,\theta}}{4 \sigma_{\phi \phi}}+\tau^2_{,\theta}}.
\end{eqnarray}
Now we can write the mean curvature in terms of the derivatives of $\rho$ and $f$ with respect to $\theta$.  The principle curvatures in the $\theta$ and $\phi$ directions are
\begin{eqnarray}
\hat{k}_{\theta \theta} &=& \frac{f_{,\theta}\rho_{,\theta \theta}-\rho_{,\theta}f_{,\theta \theta}}{(f_{,\theta}^2+\rho_{,\theta}^2)^{3/2}} \hbox{ and} \\
\hat{k}_{\phi \phi} &=& -\frac{f_{,\theta}}{\rho\sqrt{f_{,\theta}^2+\rho_{,\theta}^2}},
\end{eqnarray}
respectively.  The mean curvature is the sum of the principle curvatures and is given by
\begin{equation}
\label{eq:refQLE}
\hat{k}=-\left(\frac{{f}_{,\theta}^3+ \rho\,  \rho_{,\theta}\,  f_{,\theta \theta} +  f_{,\theta}\left(\rho_{,\theta}^2-  \rho\, \rho_{,\theta \theta} \right)}{\left(f_{,\theta}^2+\rho_{,\theta}^2 \right)^{3/2}}\right) \, \frac{1}{\rho}.
\end{equation}
We integrate Eq.~\ref{eq:refQLE} over $\hat{\mathcal{S}}$ to get the contribution to QLE from the reference action.  With Eqs~\ref{eq:term1},~\ref{eq:term2} and~\ref{eq:refQLE}, the W-Y QLE functional is completely determined by $\tau_{,\theta}$, $\tau_{,\theta \theta}$ and $\tau_{,\theta \theta \theta}$.

\section{The critical point of the W-Y QLE functional for constant radial surfaces}
\label{sec:zeromin}
In section 6 of~\cite{WY:comm},  W-Y derived the Euler-Lagrange equation of $E[\tau]$, which is given by
\begin{multline}
\label{eq:EuLa}
 \underbrace{-\left(\hat{k}\hat{\sigma}^{ab} - \hat{\sigma}^{ac}\hat{\sigma}^{bd}\hat{k}_{cd}\right)\frac{\nabla_b \nabla_a \tau}{\sqrt{1+|\nabla \tau|^2}}\ +}_{\left(1\right)}\\  \underbrace{\sigma^{ab} \nabla_a \left(\frac{\nabla_b \tau}{\sqrt{1+|\nabla \tau|^2}}\cosh{\alpha}|\vec{H}|\right)}_{\left(2\right)}-\underbrace{\Delta \alpha}_{\left(3\right)}-\underbrace{\sigma^{ab}\nabla_a \langle \vec{\zeta}_{b}\ |\ \nabla \vec{\bar{v}}\ |\ \vec{\bar{u}} \rangle}_{\left(4\right)}=0.
\end{multline}
All covariant derivatives are taken with respect to the 2-metric on $\mathcal{S}$ except for the covariant derivative on $\vec{\bar{v}}$, which is taken with respect to the spacetime metric $\pmb{g}$. To show that $\tau=0$ is a solution to Eq.~\ref{eq:EuLa}, we write each term explicitly in terms of $\tau$.

The first term of Eq.~\ref{eq:EuLa} written explicitly in terms of $\tau$ is given by
\begin{equation}
\left(1\right)=-\frac{1}{\sqrt{1+\frac{\tau^2_{,\theta}}{\sigma_{\theta \theta}}}}\left(\frac{\left(\tau_{,\theta \theta}-\Gamma^{\theta}_{\theta \theta}\tau_{,\theta}\right)\left(\hat{k}-\hat{\sigma}^{\theta \theta}\hat{k}_{\theta \theta}\right)}{\sigma_{\theta \theta}+\tau^2_{,\theta}}-\frac{\Gamma^{\theta}_{\phi \phi}\tau_{,\theta}}{\sigma_{\phi \phi}}\left(\hat{k}-\hat{\sigma}^{\phi \phi}\hat{k}_{\phi \phi}\right)\right).
 \end{equation}
 The second term is
 \begin{equation}
 \left(2\right)=\partial_{\theta} \left(\frac{|\vec{H}|\cosh{\alpha}}{\sqrt{1+\frac{\tau^2_{,\theta}}{\sigma_{\theta \theta}}}}\right)\frac{\tau_{,\theta}}{\sigma_{\theta \theta}} + \frac{|\vec{H}|\cosh{\alpha}}{\sqrt{1+\frac{\tau^2_{,\theta}}{\sigma_{\theta \theta}}}} \Delta \tau
 \end{equation}
 where $\cosh{\alpha}$ and $\Delta \tau$ are given by Eqs.~\ref{eq:cosh} and~\ref{eq:laptau}, respectively.
Term 3 is simply
\begin{equation}
\left(3\right)=\frac{1}{\sigma_{\theta \theta}}\left(\alpha_{,\theta \theta} - \Gamma^\theta_{\theta \theta} \alpha_{,\theta} \right) - \frac{\Gamma^\theta_{\phi \phi}\alpha_{,\theta}}{\sigma_{\phi \phi}}.
\end{equation}
Let 
\begin{equation}
\label{eq:va}
V_a=\langle \vec{\zeta}_a\ |\ \nabla \vec{\bar{v}}\ |\ \vec{\bar{u}}\rangle,
\end{equation}
the last term in Eq.~\ref{eq:EuLa} is given by
\begin{equation}
\label{eq:term4}
\left(4\right)=\sigma^{ab}\nabla_a V_b=\frac{1}{\sigma_{\theta \theta}}\left(V_{\theta,\theta}-\Gamma^{\theta}_{\theta \theta}V_{\theta}\right)+\frac{1}{\sigma_{\phi \phi}}\left(V_{\phi,\phi}-\Gamma^{\theta}_{\phi \phi}V_\theta\right).
\end{equation}
It is easy to see that the first three terms vanish for $\tau=0$.  Next we show that the fourth term also vanishes for $\tau=0$.  

Writing $V_\theta$ and $V_\phi$, one gets
\begin{eqnarray}
\label{eq:vtheta}
V_\theta&=&\partial_\theta \bar{v}^\nu \bar{u}_\nu + \Gamma_{\theta \alpha}^\nu\bar{v}^\alpha \bar{u}_\nu\ \hbox{and} \\
\label{eq:vphi}
V_\phi&=&\Gamma_{\phi r}^t \bar{v}^r \bar{u}_t.
\end{eqnarray}
From Eqs.~\ref{eq:utilde} and~\ref{eq:vtilde}, it is clear that $\{\vec{\bar{u}},\vec{\bar{v}}\}=\{\vec{u'},\vec{v'}\}$ for $\tau=0$. It is also clear that the first term of Eq.~\ref{eq:vtheta} is equal to zero since $\vec{v'}$ only has a radial component and the contravariant components of $\vec{u'}$ are only non-zero for time.  Furthermore, the second term of Eq.~\ref{eq:vtheta} reduces to $\Gamma_{\theta r}^t v'^r u'_t$ where $\Gamma_{\theta r}^t=0$. This gives $V_\theta=0$.  Inserting $V_\theta=0$ into Eq.~\ref{eq:term4} gives $\left(4\right)=V_{\phi,\phi}/\sigma_{\phi \phi}$.  Since $V_\phi$ is independent of $\phi$, term 4 vanishes.  This shows explicitly that $\tau=0$ is a critical point regardless of one's choice of $R$.   Indeed, it was shown in \cite{cyw} that for any axi-symmetric surface, the fourth term of Eq.~\ref{eq:EuLa} always vanishes and $\tau=0$ is always a solution. However, $\tau=0$ is not necessarily a local or global minimum, see \cite{cw, cwy} for a criterion for local minimum of a critical point in terms of a mean curvature inequality. 

\section{Numerical Results}
\label{sec:numerical}
In this section, we apply the direct search algorithm developed by Torczon~\cite{torczon:diss} to minimize $E[\tau\left(\theta\right)]$, which is given by Eqs.~\ref{eq:term1},~\ref{eq:term2} and~\ref{eq:refQLE}, in the space of coefficients.  Without loss of generality, we will use $a=M=1$ for our numerical analysis unless stated otherwise.  The value of $r^*$ is approximately $1.65$ for this choice of $a$ and $M$.  

\begin{figure}[H]
\centering
\includegraphics[width=5in]{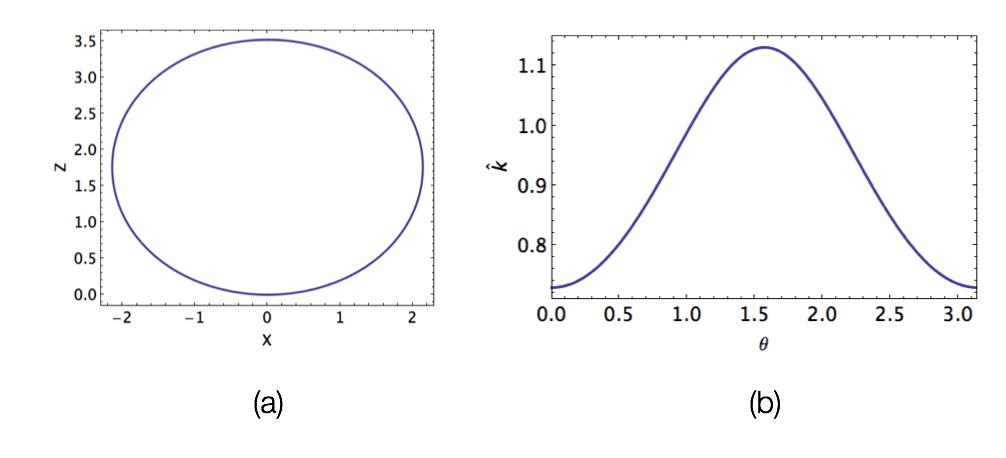}
\caption{Figure (a) gives a $\phi=0$ cross section of the convex shadow defined by Eq.~\ref{eq:fo} at $R=3/2$.  The mean curvature of the that cross section is given in Figure (b).  Notice that it is well behaved at the poles, $\theta=0,\pi$.}
\label{fig:shadow}
\end{figure}

To apply the direct search algorithm, we use a Fourier expansion to express $\tau_{,\theta}$ as
\begin{equation} 
\label{eq:tau}
\tau_{,\theta}\left( \theta \right) = F_0\left( \theta \right) + \sum^\kappa_{n=1}{a_n\sin{n\theta}}
\end{equation}
where $\theta$ is the polar angle in Boyer-Lindquist coordinates, $F_0\left(\theta\right)$ is an initial guess of the optimal $\tau_{,\theta}$ and $a_n$ are the Fourier coefficients.  Symmetry about the equator excludes all but the odd values of the Fourier coefficients of $\sin(n\theta)$.  The expansion lacks cosine modes due to boundary conditions on the derivative of $\tau$ at the poles.  We choose our initial guess to be
\begin{equation}
\label{eq:fo}
F_0\left(\theta\right)=\sqrt{\frac{\sigma_{\phi \phi}}{\sin^2{\theta}} - \sigma_{\theta \theta}} .
\end{equation}
This function gives an integrand of $E[\tau\left(\theta\right)]$ that is well behaved at the poles.  It also gives an initial guess reasonably close to a solution of the Euler-Lagrange equation for all radii.  The image of the convex shadow and its mean curvature at $R=3/2$  are shown in Fig.~\ref{fig:shadow}.
\noindent There is nothing notable about $R=3/2$;  we simply use this for as an illustrative example for surfaces with $R < r^*$.  We will continue to use this radius for further examples.  All statements made for $R=3/2$ apply equally to all radii below $r^*$ unless specified otherwise.

The complexity of the space of coefficients increases with the dimension.  As one increases the number of coefficients used to minimize $E[\tau\left(\theta\right)]$, the likelihood of getting caught in local minima increases.  To mitigate this difficulty, we begin with just one Fourier coefficient set to zero.  We then apply the direct search algorithm to find the smallest real value of $E[\tau\left(\theta\right)]$ in the space of $a_1$.  Once $a_1$ is obtained, we add $a_3=0$ and search in the space of $a_1$ and $a_3$.  Here we allow both $a_1$ and $a_3$ to change until we find the minimum in two dimensions.  We iteratively increase the number of coefficients until the change in $E_{min}$ is at least less than $10^{-2}$ for each additional coefficient added.  The number of coefficients needed increases as one approaches $r_+$ due to increasing curvature gradients of $\mathcal{S}$.  Our direct search algorithm was coded using Mathematica.  All integrals were done using the NIntegrate function.   

\subsection{The boundary separating admissible and non-admissible values of the W-Y QLE functional}
\label{sec:boundary}
There are three criteria within the W-Y QLE formalism that determine whether a choice of $\tau$ is admissible.  These criteria can be found in section 4 of~\cite{WY:physrev} as well as section 5 of~\cite{WY:comm}.  The purpose of the second and third criteria is to ensure that the value of $E[\tau\left(\theta\right)]$ is positive.  We will not focus on these since we do not obtain negative energies for any of our results.  Instead, we will focus on the first criteria, which is
\begin{equation}
\label{eq:violate}
{\mathcal K} - \left(1+|\nabla \tau |^2\right)^{-1} det\left( \nabla_a\nabla_b \tau \right) > 0
\end{equation}  
where $\mathcal{K}$ is the Gaussian curvature of $\mathcal{S}$ and all covariant derivatives are taken with respect to $\pmb{\sigma}$.  This requires that the Gaussian curvature of the convex shadow given a choice of $\tau$ is strictly positive everywhere. If this criteria is met, W-Y can guarantee the existence and uniqueness of $E[\tau\left(\theta\right)]$.  Unfortunately, our analysis indicates that this criteria can not be met at $\tau=0$. 

\begin{figure}[H]
\centering
\includegraphics[width=4in]{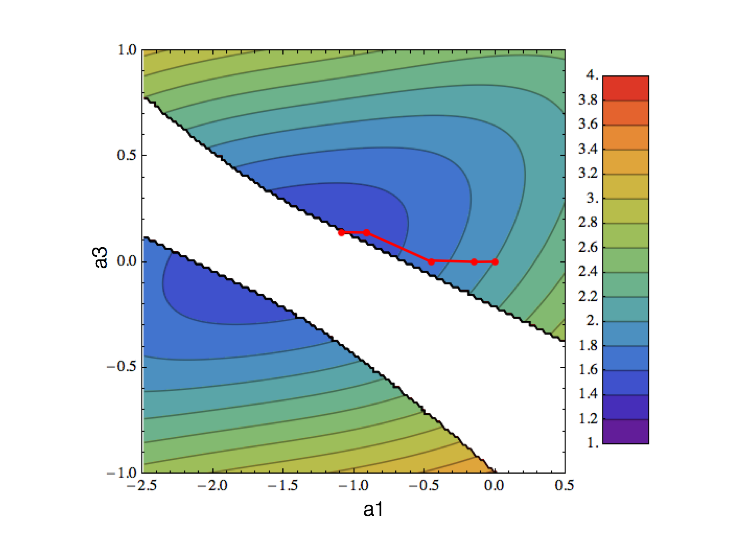}
\caption{This figure is a contour plot of $E[\tau\left(\theta\right)]$ as a function of the two Fourier coefficients with $a_1$ on the x-axis and $a_3$ on the y-axis.  The line connecting the points indicates the path taken by the simplex method when optimizing the functional using just two coefficients initialized at zero.  The point furtherest to the right is the initial guess while the point on the boundary is $E_{min}$.  Our numerical results are consistent with the smallest real value occurring on the boundary separating real and complex energies.} 
\label{fig:phase}
\end{figure}

While minimizing in the space of coefficients, we discovered a boundary separating $\tau$'s with real values of $E[\tau\left(\theta\right)]$ from those with complex values.  This can be seen in Fig.~\ref{fig:phase} where we use two Fourier coefficients, $a_1$ and $a_3$, to visualize the QLE landscape.
\noindent The white gap in the middle of the plot represents complex values of $E[\tau\left(\theta\right)]$ whose existence can be understood by examining Eq.~\ref{eq:df}.  Here, complex energies arise for choices of $\tau_{,\theta}$ that satisfy
\begin{equation}
\label{eq:imag}
\tau^2_{,\theta} < \frac{\sigma^2_{\phi \phi,\theta}}{4\sigma_{\phi \phi}} - \sigma_{\theta \theta}.
\end{equation}    
We will show that these choices of $\tau$ are inadmissible using our numerical results.

\begin{figure}[H]
\centering
\includegraphics[width=5in]{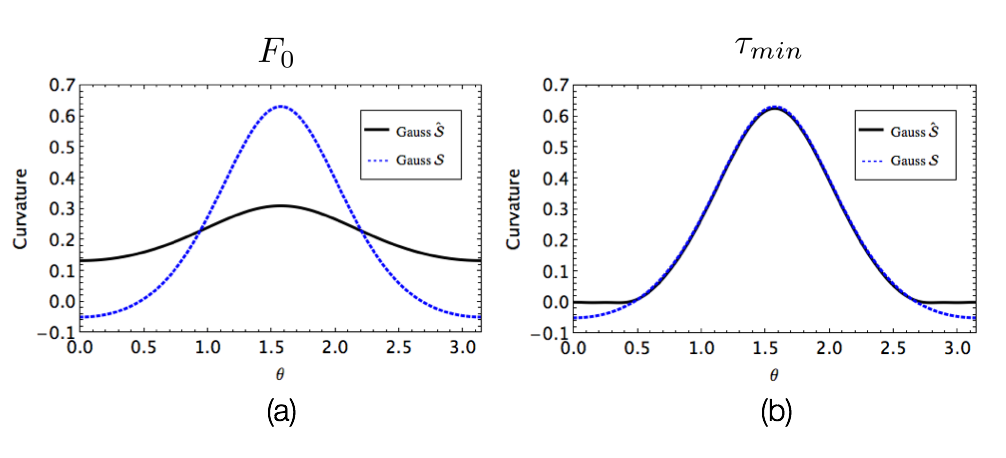}
\caption{Figure (a) compares the Gaussian curvature between the convex shadow for the initial guess and $\mathcal{S}$.  Figure (b) compares the Gaussian curvature between the convex shadow for $\tau_{min}$ and $\mathcal{S}$. }
\label{fig:gausshat}
\end{figure}

To demonstrate that choices of $\tau$ with complex energies are not admissible, we compare the Gaussian curvature of $\mathcal{S}$ and $\hat{\mathcal{S}}$ for the initial guess and $\tau_{min}$ in Fig.~\ref{fig:gausshat}.
\noindent From Fig.~\ref{fig:gausshat}a, we see that the Gaussian curvature of the convex shadow for the initial guess is strictly positive and significantly different than the curvature of $\mathcal{S}$.  On the other hand, Fig.~\ref{fig:gausshat}b shows that the Gaussian curvature of the convex shadow at $\tau_{min}$ is similar to the curvature of $\mathcal{S}$ within the interval of positive Gaussian curvature.  Outside of this interval, the Gaussian curvature of $\mathcal{S}$ becomes negative while the shadow's curvature is flat.  This indicates that the optimization algorithm tends toward a $\tau$ that embeds $\mathcal{S}$ into $\mathbb{R}^3$ as much as possible.  In fact, if we allow the algorithm to cross the boundary of admissible solutions by taking the real part of the QLE functional, it converges to $\tau=0$.  This implies that choices of $\tau$ within the boundary do not have shadows with strictly positive Gaussian curvature.  We believe this is due to the unnecessary restriction that $\tau$ is a function of $\theta$ only. In general, we should allow $\tau$ to be dependent of both $\theta$ and $\phi$ and solve both the isometric embedding equation and the Euler-Lagrange equation.

\subsection{The physical relevance of $E_{min}$}
\label{sec:physrel}
\begin{figure}[H]
\centering
\includegraphics[width=4in]{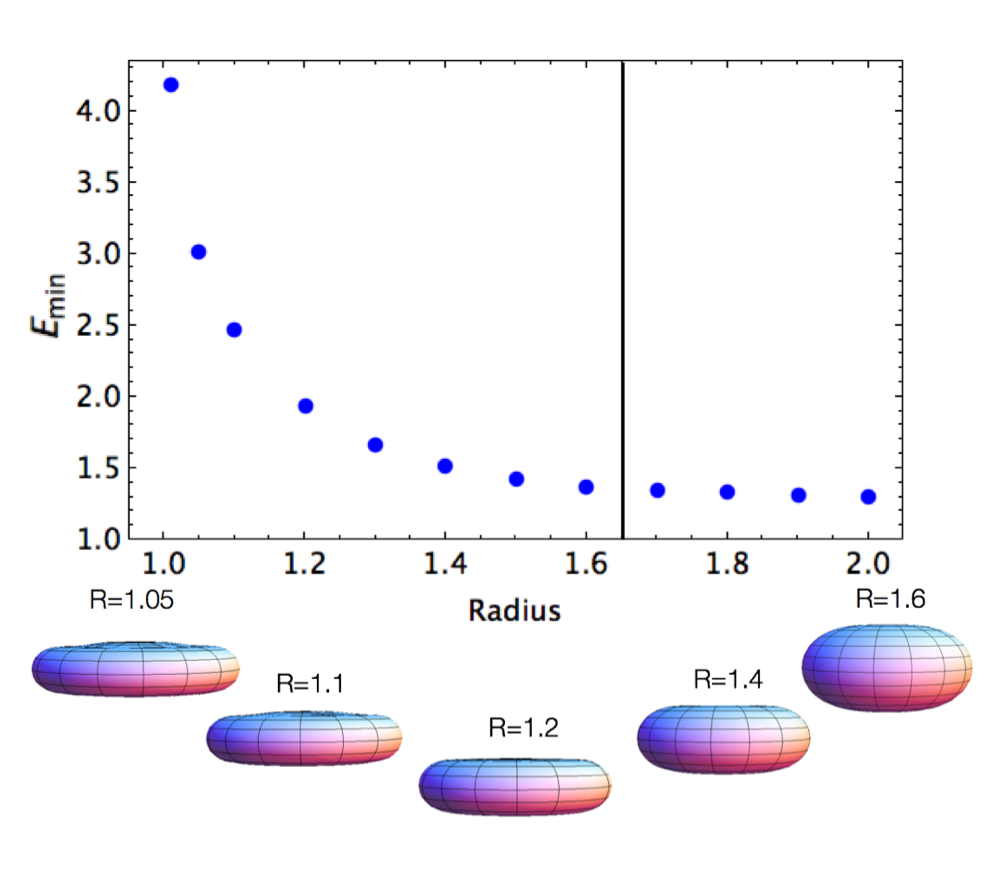}
\caption{In this figure we plot $E_{min}$ as a function of radius where $r_+=1$ is the event horizon.  The vertical dividing line is located at the critical radius $r^*\approx1.65$.  Below $r^*$ we plot the smallest real value of $E[\tau\left(\theta\right)]$ while above $r^*$ we plot the B-Y QLE.  These two values agree above $r^*$.  We show the evolution for the convex shadow by plotting it for $R=\{1.05,1.1,1.2,1.4,1.6\}$.  Notice how they become more flat as one approaches the event horizon.}
\label{fig:evo}
\end{figure}

\begin{figure}[H]
\centering
\includegraphics[width=5.2in]{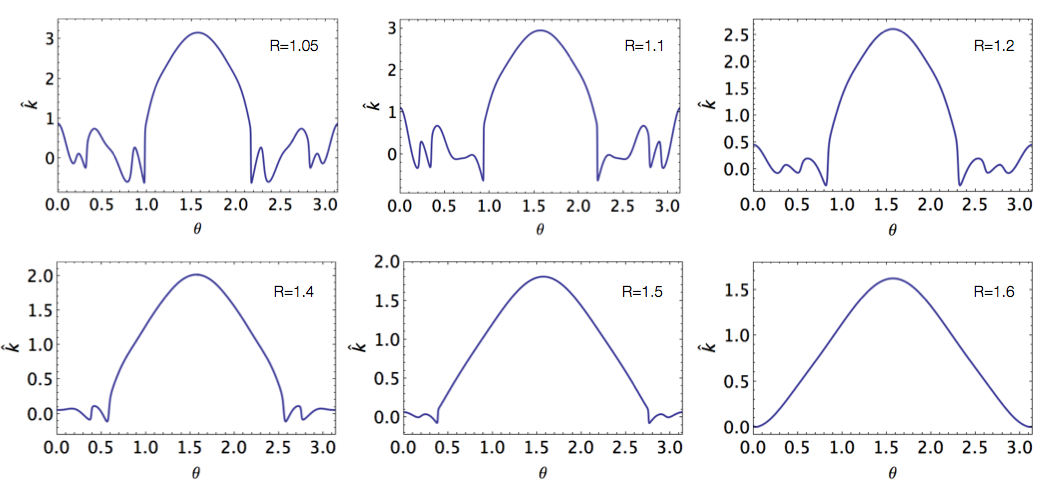}
\caption{Here are the mean curvature of the shadows at $\tau_{min}$ for several values of $R$.  For each plot, there is a distinct change in behavior approximately at $\theta=\theta^*$ and $\theta=\pi-\theta^*$, where $\theta^*$ is the angle where the Gaussian curvature of $\mathcal{S}$ becomes negative.  This is due to the Gibbs' phenomena that emerges from the sudden drop to a flat function.}
\label{fig:meancurvs}
\end{figure}

In this section we analyze the physical behavior of $E_{min}$ and compare it to what one would expect based on the results of Martinez.  We begin by plotting $E_{min}$ as a function of $R$ in Fig.~\ref{fig:evo}.  
\noindent The vertical dividing line is placed at the critical radius $r^*\approx1.65$.  Above $r^*$, $E_{min}$ is equivalent to the B-Y QLE.  Below the plot in Fig.~\ref{fig:evo} are the convex shadows at $\tau_{min}$ associated with radii $R=\{1.05, 1.1, 1.2, 1.4, 1.6\}$.  The mean curvature of these shadows can be see in Fig.~\ref{fig:meancurvs}.
\noindent These plots show a Gibbs' phenomena that occurs when the Gaussian curvature of $\mathcal{S}$ becomes negative, which, as we showed in Fig.~\ref{fig:gausshat}, is also when the Gaussian curvature of $\hat{\mathcal{S}}$ at $\tau_{min}$ becomes zero.  To physically interpret the results in Fig.~\ref{fig:evo},  we analyze $E_{min}$ at the outer event horizon for black holes with increasing angular momentum.  This will give us a reference on how $E_{min}$ should behave at the event horizon once the angular momentum exceeds $a=\sqrt{3}M/2$.  We will also interpret the results by looking at the field of observers associated with $E_{min}$ and compare them to the Eulerian observers chosen by B-Y.

\begin{figure}[H]
\centering
\includegraphics[width=5.1in]{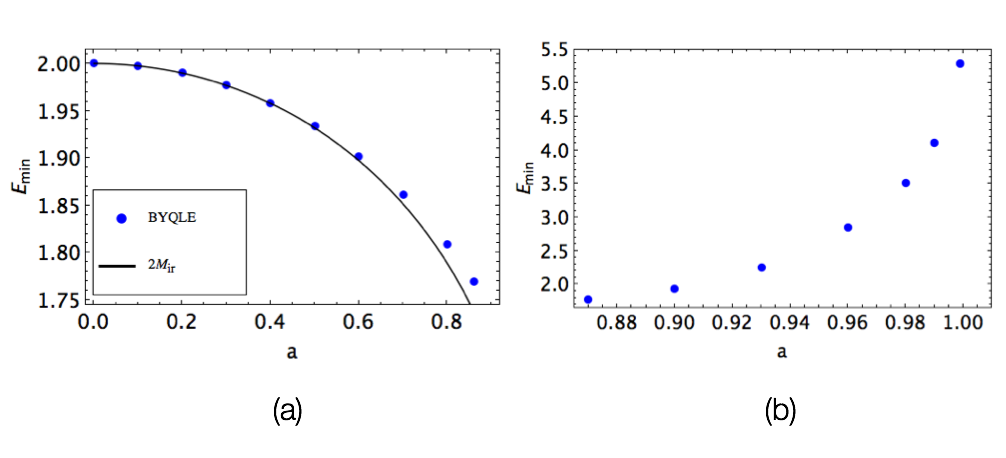}
\caption{In figure a we have the B-Y QLE at $r_+$ for angular momentum less than $\sqrt{3}M/2=0.866$.  We also plot twice the irreducible mass as predicted by Martinez for black holes with $a/R<<1$.  In figure b, we have $E_{min}$ at an $\epsilon$ above $r_+$ for black holes with angular momentum ranging from  0.87 to .999.}
\label{fig:angular}
\end{figure}

In Fig.~\ref{fig:angular}a, we plot $E_{min}$, which is equivalent to the B-Y QLE for $a<\sqrt{3}M/2$, at various values of $a$ between $\left(r_+,0\right)$ and $\left(r_+,\sqrt{3}M/2\right)$. 
\noindent We see that these two plots agree for $a \leq 0.4$.  As the angular momentum grows, the low angular momentum approximation starts to deviate from the B-Y QLE.  The most important feature of this plot is the fact that the B-Y QLE decreases as angular momentum increases.  Looking at Fig.~\ref{fig:angular}b, we plot $E_{min}$ at $r_+ + \epsilon$, where $\epsilon=10^{-5}$, for values of $a$ between points $\left(r_+,\sqrt{3}M/2\right)$ and $\left(r_+,1\right)$.  Here we see that the $E_{min}$ predicts a growing energy with increased angular momentum.  So why is $E_{min}$ significantly greater than the predicted $2M_{ir}$ and why does it disagree with the trend of decreasing QLE with increased angular momentum?  The reason is due to the field of observers that are chosen at $\tau_{min}$.

\begin{figure}[H]
\centering
\includegraphics[width=3in]{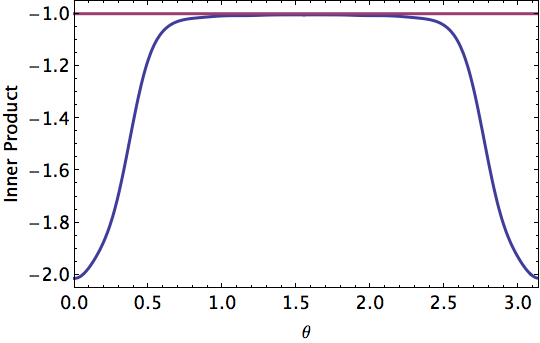}
\caption{In this plot we have the inner product between the observers chosen by W-Y at $\tau_{min}$ and the Eulerian observers that would be chosen by B-Y.  The horizontal line at -1.0 is what the inner product would be if \noindent the observers agreed for all $\theta$.  The curved line is the actual inner product between the two observer fields.  From here it is clear that the two agree within some $\epsilon$ difference between $0.7<\theta<\pi-0.7$.  The values of $\theta$ where they do not agree are outside the interval of embeddability.}
\label{fig:inner}
\end{figure}

Assuming the isometric embedding of $\mathcal{S}$ is a surface of revolution, the interval for which $\mathcal{S}$ is embeddable in $\mathbb{R}^3$ is determined by Eq.~\ref{eq:df} with $\tau=0$.  For surfaces in the regime of strictly positive Gaussian curvature, $\sigma_{\theta \theta}$ is strictly greater than or equal to $\frac{\sigma^2_{\phi \phi,\theta}}{4 \sigma_{\phi \phi}}$.  Surfaces with regions of negative Gaussian curvature can only be partially embedded in $\mathbb{R}^3$ between $\theta^{**} <\theta <\pi-\theta^{**}$. Here, $\theta^{**}$ is the smaller root of Eq.~\ref{eq:df} with $\tau=0$.  We will refer to this interval as the ``interval of embeddability".  In Fig.~\ref{fig:inner}, we plot the inner product between the field of observers given by W-Y at $\tau_{min}$ with the Eulerian observers that would be chosen by B-Y as a function of $\theta$. 
\noindent We see that the observers at $\tau_{min}$ and the Eulerian observers agree within some $\epsilon$ around $0.7<\theta<\pi-0.7$.  For this choice of angular momentum and radius, $\theta^{**}$ is approximately equal to 0.64.  This shows that $E_{min}$ chooses the Eulerian observers within the interval of embeddability and smoothly transitions to observers that are boosted with respect to the Eulerian observers outside of this interval.  

As one approaches $\left(r_+,M\right)$, the interval of embeddability decreases.  This implies that more observers chosen by the W-Y QLE procedure at $\tau_{min}$ are boosted with respect to the Eulerian observers. We also found that the magnitude of the boosts increases as one approaches $\left(r_+,M\right)$.  This is why $E_{min}$ has a growing energy with increased angular momentum.  It also explains why the $E_{min}$ at the event horizon is significantly greater than twice the irreducible mass when $a>\sqrt{3}M/2$. 
\section{Conclusion and Further Discussion}
\label{sec:conclusion}
In this paper, we analyzed the W-Y QLE functional with the restriction that $\tau$ is only a function of $\theta$ for constant radial surfaces with $R<r^*$.  These surfaces may not be embeddable in $\mathbb{R}^3$, but they are embeddable in $\mathbb{R}^{3,1}$.   We discovered an open region of complex values for $E[\tau\left(\theta\right)]$ while minimizing the functional in the space of coefficients.  Our results suggest that the smallest real value of $E[\tau\left(\theta\right)]$ lies on the boundary separating real and imaginary energies. Our results also suggest that there does not exist a convex shadow whose Gaussian curvature is strictly non-negative for choices of $\tau$ within the region of complex energies.   

We also analyzed the behavior of $E_{min}$ to gleam some insight on its possible physical relevance.  In Fig.~\ref{fig:evo}, we saw a sudden increase in $E_{min}$ for surfaces with $R<r^*\approx1.65$.  It is uncertain if these energies are physically meaningful since no results exists for such surfaces.  To gain some clarity, we examined $E_{min}$ at the event horizon as a function of angular momentum.  For $a<\sqrt{3}M/2$, the results of Martinez suggest that the QLE is comparable to the irreducible mass of the black hole, which decreases with increasing angular momentum.  Above $\sqrt{3}M/2$, $E_{min}$ increases with increasing angular momentum.  We attributed this change in behavior to the difference between the Eulerian observers chosen by B-Y  and the field of observers chosen by W-Y at $\tau_{min}$.  In Fig.~\ref{fig:inner}, we showed that the W-Y observers at $\tau_{min}$ agree with the Eulerian observers within the interval of embeddability and transition to boosted observers outside of this interval.  Our results are contingent on $\tau$ being a function of $\theta$ alone.  For a true understanding of the W-Y QLE applied to extreme Kerr spacetimes near the event horizon, one must allow $\tau$ to be a function of both $\theta$ and $\phi$.  This is an interesting avenue for future research.

\section*{Acknowledgements}
We wish to thank Po-Ning Chen, Rory Conboye, Matthew Corne, and Ye-Kai Wang for stimulating discussions. We also thank the the Information Directorate of the Air Force Research
  Laboratory and the Griffiss Institute for providing us with an
  excellent environment for research. This work was supported in part
  through the VFRP and SFFP program, as well as AFRL grant
  \#FA8750-15-2-0047. 
 
\appendix
\subsection{Physical motivations behind the W-Y QLE formalism}
\label{sec:app}
Let $\mathcal{M}$ be an arbitrary spacetime manifold.  The B-Y QLE energy given by Eq.~\ref{eq:qleBY} does not give a general description on how to choose the field of observers in $\mathcal{M}$, nor does it give the reference space.  For stationary spacetimes, B-Y suggested the Eulerian observers associated with maximal hypersurfaces as their observers and $\mathbb{R}^3$ as their reference space.  This choice is reasonable when the extrinsic curvature of $\mathcal{S}$ along $\vec{u}$ vanishes and the embedding of $\mathcal{S}$ exists.  However, it was shown that these choices do not work for surfaces in general and can give non-zero values of QLE for flat spacetimes~\cite{Neil:physrev}.  This is due to the second term in Eq.~\ref{eq:qleBY}.  If the extrinsic curvature of $\mathcal{S}_t$ along $\vec{u}$ does not vanish, there is no way to account for this curvature in $\mathbb{R}^3$ when computing the reference energy.  To address this problem, W-Y used the flat spacetime as their reference space.  This extends the application of the B-Y QLE to dynamical spacetimes.

The extension of the reference space from $\mathbb{R}^3$ to Minkowski space $\mathbb{R}^{3,1}$ creates a new challenge. Since $\mathcal{S}_t$ is a co-dimension two surface with respect to $\mathbb{R}^{3,1}$, the isometric embedding equations are underdetermined. To address this problem, W-Y introduced the scalar field $\tau$ on $\mathcal{S}_t$ which allows them to define a unique embedding into Minkowski space up to a choice of $\tau$.  They then construct a procedure to associate each choice of $\tau$ with two observer fields which are used to compute QLE.  One field exists in the physical space and is denoted as $\vec{t}=N\vec{\bar{u}}+\vec{N}$ while the other exists in Minkowski space and is denoted as $\vec{t}_0=N\vec{u}_0+\vec{N}$. These observers are chosen such that the extrinsic curvature of $\mathcal{S}_t$ along $\vec{\bar{u}}$ embedded in $\mathcal{M}$ is equal to the extrinsic curvature of $\mathcal{S}_t$ along $\vec{u}_0$ embedded in $\mathbb{R}^{3,1}$.  The notation $\vec{\bar{u}}$ is used to distinguish the unique timelike normal on $\mathcal{S}_t$ whose extrinsic curvature agrees with $\vec{u}_0$ as opposed to an arbitrary timelike normal $\vec{u}$.  This matching of extrinsic curvature along timelike normals is given by the constraint
\begin{equation}
\label{eq:junction}
\langle \vec{\bar{u}},\vec{H} \rangle = \langle \vec{u}_0,\vec{H}_0 \rangle
\end{equation}
where $\vec{H}_0$ is the mean curvature vector of $\mathcal{S}_t$ embedded in $\mathbb{R}^{3,1}$.
This addresses the problem of the second term in Eq.~\ref{eq:qleBY}.  Next we discuss the isometric embedding into Minkowski space and how the lapse and shift are chosen.  

Let $i:\mathcal{S}_t \hookrightarrow \mathbb{R}^{3,1}$ represent an isometric embedding of $\mathcal{S}_t$ into Minkowski space.  In principle, one would compute the reference energy using a field of observers who are at rest with respect to $i\left(\mathcal{S}_t\right)$.  If we work in the rest frame of these observers, at each point $p \in i\left(\mathcal{S}_t\right)$ we have $\vec{t}_0 = \{1,0,0,0\}$.  Let $\tau$ be the time component of $i\left(\mathcal{S}_t\right)$,  then the embedding takes the form $\vec{x}_0=\{\tau,x^1,x^2,x^3\}$.  One can alternatively write the embedding as
\begin{equation}
\label{eq:embed}
\vec{x}_0 = \vec{\hat{x}} + \tau \vec{t}_0
\end{equation} 
where $\vec{\hat{x}}=\{0,x^1,x^2,x^3\}$ are the spatial coordinates of $i \left(\mathcal{S}_t\right)$ that lie in a three dimensional Euclidean plane orthogonal to $\vec{t}_0$.  This projection $\hat{\mathcal{S}}$ onto $\mathbb{R}^3$ is defined as the shadow of $i \left(\mathcal{S}_t\right)$ with respect to $\vec{t}_0$.  Vectors with hats exist on the shadow, while vectors with the zero subscript exist on $i \left(\mathcal{S}_t\right)$.  
\begin{figure}[h]
\centering
\includegraphics[width=3in]{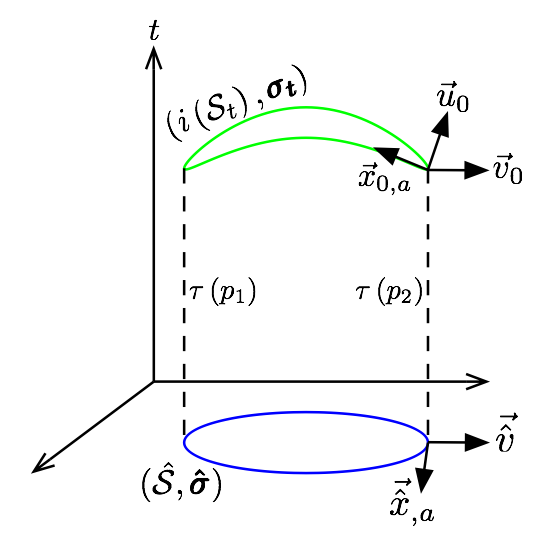}
\caption{This figure has one spatial dimension suppressed.  The procedure for isometrically embedding $\mathcal{S}_t$ into $\mathbb{R}^{3,1}$ is as follows.  First isometrically embed $\hat{\mathcal{S}}$ into $\mathbb{R}^3$.  If $\pmb{\hat{\sigma}}$ has strictly positive Gaussian curvature, this embedding is guaranteed to exist via the Nirenberg and Pogorelov embedding theorem. Next, extend each point $p \in \hat{\mathcal{S}}$ along $\vec{t}_0$ by defining the time coordinate as $\tau\left(p\right)$.  This extension into $\mathbb{R}^{3,1}$ is showed by the dotted lines at points $p_1$ and $p_2$.  To find the unique normal basis $\{\vec{u}_0,\vec{v}_0\}$, one begins by computing the tangent vectors $\vec{\hat{x}}_{,a}$ and normal vector $\vec{\hat{v}}$ of $\hat{\mathcal{S}}$.  The tangent vectors $\vec{x}_{0,a}$ of $i \left(\mathcal{S}_t\right)$ have the same spatial components as $\vec{\hat{x}}_{,a}$ but their temporal components are $\tau_{,a}$.  W-Y choose the spatial components of the spacelike normal $\vec{v}_0$ to be identical  to $\vec{\hat{v}}$, each with a zero  temporal component.  Finally, $\vec{u}_0$ is given by $\left(\vec{x}_{0,1}\wedge \vec{x}_{0,2}\wedge\vec{v}_0\right)^*$.   
}
\label{referenceEmbedding}
\end{figure}
Starting from Eq.~\ref{eq:embed}, the metric of the shadow $\pmb{\hat{\sigma}}$ is given by
\begin{eqnarray}
\label{eq:proof1}
\langle\vec{x}_{,a}\ |\ \vec{x}_{,b}\rangle &=&\langle \vec{\hat{x}}_{,a}+\tau_{,a} \vec{t}_0\ |\ \vec{\hat{x}}_{,b}+\tau_{,b} \vec{t}_0 \rangle \\
\label{eq:proof2}
&=& \langle\vec{\hat{x}}_{,a}\ |\ \vec{\hat{x}}_{,b}\rangle -\tau_{,a}\tau_{,b},
\end{eqnarray}
which implies
\begin{equation}
\label{eq:proof3}
\hat{\sigma}_{ab} = \left(\sigma_t\right)_{ab} + \tau_{,a} \tau_{,b}.
\end{equation}
The isometric embedding of $\mathcal{S}_t$ into $\mathbb{R}^{3,1}$ using $\hat{\mathcal{S}}$ and $\tau$ is shown in Fig.~\ref{referenceEmbedding}.  

A necessary condition for choosing $\tau$ requires the shadow $\hat{\mathcal{S}}$ to be a smooth convex surface in $\mathbb{R}^3$.  This condition is used to prove the existence and uniqueness of $i\left(\mathcal{S}_t\right)$ given the observer field $\vec{t}_0$.  It can be seen from the embedding theorem of Nirenberg and Pogorelov and Eq.~\ref{eq:embed} that any isometric embeddings of $\mathcal{S}_t$ in Minkowski space with the same convex shadow and scalar field $\tau$ must be congruent.  This completes the discussion on embedding $\mathcal{S}_t$ into $\mathbb{R}^{3,1}$. 

Since the field of observers in Minkowski is defined as being at rest with respect to $i\left(\mathcal{S}_t\right)$, the lapse and shift are chosen such that
\begin{equation}
\vec{t}_0=N\vec{u}_0 + \vec{N}.
\end{equation}
Using the embedding as described in Fig.~\ref{referenceEmbedding}, it can be shown that
\begin{equation}
\label{eq:lapshift}
\vec{t}_0=\{1,0,0,0\}=\sqrt{1+|\nabla \tau|^2}\ \vec{u}_0 - \nabla \tau
\end{equation}
where
\begin{equation}
\label{eq:gradtau}
\nabla \tau = \sigma_t^{ab}\tau_{,a} \vec{x}_{0,b}.
\end{equation}
Here we see that $N=\sqrt{1+|\nabla \tau|^2}$ and $\vec{N}= -\nabla \tau$.  The corresponding field of observers in $\mathcal{M}$ is
\begin{equation}
\vec{t}=\sqrt{1+|\nabla \tau|^2}\ \vec{\bar{u}} - \nabla \tau
\end{equation}
where the coordinates $\vec{x}_0$ in Eq.~\ref{eq:gradtau} are replaced with the coordinates of $\mathcal{S}_t$ in $\mathcal{M}$.  With the observer fields and the isometric embedding into $\mathbb{R}^{3,1}$ written in terms of $\tau$, the discussion on the physical motivations behind the W-Y formalism is complete. 

\subsection{A discussion on the isometric embedding theorem}
\label{sec:iso}

There are several common misconceptions about isometric embedding of a closed surface into $\mathbb{R}^3$.  We take this opportunity to address these issues. 

\vspace{\baselineskip}

\noindent 1. Isometric embeddings do not preserve symmetry: 
One reason why the current formalism does not work is because of the assumption that $\tau$ is a function of $\theta$ only, or $\tau$ is axi-symmetric. The Killing field of a Riemannian metric does not extend to the embedding, or does not extend to be a Killing field of the ambient space. In particular, it is possible that an axi-symmetric metric admits an isometric embedding into $\mathbb{R}^3$ that is not a surface of revolution. 

\vspace{\baselineskip}

\noindent2. Non-embeddability: 
The surface isometric embedding theorem guarantees the existence and uniqueness of a global isometric embedding if the Gaussian curvature is positive everywhere. However, there does not seem to be any non-trivial non-embeddability theorem. In particular, for a surface with a metric that has negative Gauss curvature at some point, isometric embedding into $\mathbb{R}^3$ is still possible. There are many closed surfaces in $\mathbb{R}^3$ with negative Gauss curvature somewhere, but these isometric embeddings are not expected to be unique. 

\vspace{\baselineskip}

\noindent 3. Global isometric embedding vs.\ local isometric embedding: 
The theorem of Frolov on the non-embeddability near a point of negative Gauss curvature~\cite{Frolov:2006} seems to contradict a well-known local isometric embedding theorem \cite{ja} that states if a surface has negative Gauss curvature at a point, then there exists a neighborhood near the point that can be isometrically embedded into $\mathbb{R}^3$. This is the local isometric embedding theorem which holds as long as the Gauss curvature is positive, negative, or changes sign cleanly. This violation implies that Frolov's theorem does not necessarily eliminate the existence of embeddings into $\mathbb{R}^3$ for these surfaces.  In particular, one can not rule out embeddings that are not surfaces of revolution.

\bibliographystyle{unsrt}
\bibliography{qle.3.bib} {}

\end{document}